# ROSA: Robust sparse adaptive channel estimation in the presence of impulsive noises

Guan Gui[1], Li Xu[1], and Nobuhiro Shimoi[2]

1. Department of Electronics and Information Systems, Akita Prefectural University, Yurihonjo, 015-0055 Japan. E-mails: {guiguan, xuli}@akita-pu.ac.jp

2. Department of Machine Intelligence and Systems Engineering, Akita Prefectural University, Yurihonjo, 015-0055, Japan. E-mail: shimoi@akita-pu.ac.jp

**Abstract**

Based on the assumption of Gaussian noise model, conventional adaptive filtering algorithms for reconstruction sparse channels were proposed to take advantage of channel sparsity due to the fact that broadband wireless channels usually have the sparse nature. However, state-of-the-art algorithms are vulnerable to deteriorate under the assumption of non-Gaussian noise models (e.g., impulsive noise) which often exist in many advanced communications systems. In this paper, we study the problem of RObust Sparse Adaptive channel estimation (ROSA) in the environment of impulsive noises using variable step-size affine projection sign algorithm (VSS-APSA). Specifically, standard VSS-APSA algorithm is briefly reviewed and three sparse VSS-APSA algorithms are proposed to take advantage of channel sparsity with different sparse constraints. To fairly evaluate the performance of these proposed algorithms, alpha-stable noise is considered to approximately model the realistic impulsive noise environments. Simulation results show that the proposed algorithms can achieve better performance than standard VSS-APSA algorithm in different impulsive environments.

# 1. Introduction

Broadband transmission is becoming more and more important in advanced wireless communications systems [1]–[3]. The main impairments to the systems are due to multipath fading propagation as well as additive noise interferences. Hence, accurate channel state information (CSI) is required for coherence detection [1]. Based on the assumption of Gaussian noise model, channel estimation has been extensively studies in the literatures [4]–[8]. For example, to solve the problem of iterative interference cancellation in orthogonal frequency division multiplexing (OFDM) systems using training sequence as the guard interval, [4] proposed a smart solution using a small part of interference-free region within the training sequence to reconstruct the high-dimensional channel under the framework of compressive sensing, where interference cancellation can be completely avoided. Another elegant solution is that the temporal correlation of wireless channels can be integrated into the sparse channel model to generate multiple sparse channel vectors to be simultaneously reconstructed with further improved performance [5]. A well-known example is that sparse channel estimation can be also applied to the emerging massive multiple-input multiple-output (MIMO) to reduce the required channel training overhead [6],[7]. In addition, adaptive filter theory based second-order statistics based least mean square (SOS-LMS) algorithm has been widely used to estimate channels due to its simplicity and robustness [8]. However, the performance of SOS-LMS is vulnerable to deteriorate by impulsive interferences in aforementioned communications systems [9]. Such impulsive noise, which results from natural or man-made electromagnetic waves, usually has a long tail distribution and violates the commonly used Gaussian noise assumption [10]. To intuitively illustrate the distribution of impulsive noise, we consider an alpha-stable noise model which is used to describe impulsive interferences [9]. As a typical example, Fig. 1 demonstrates the distribution differences between impulsive and Gaussian noises.

To mitigate the harmful interferences, it is necessary to develop robust channel estimation algorithms. Based on the assumption of dense finite impulse response (FIR), recently, several effective adaptive channel estimation algorithms have been proposed to achieve the robustness against impulsive interferences [11]–[13]. In [11], variable step-size (VSS) sign algorithm based adaptive channel estimation was proposed to achieve performance gain. In [12], an standard VSS affine projection sign algorithm (VSS-APSA) was proposed and its improved version was also proposed in [13]. However, FIR of real wireless channel is often modeled as sparse or cluster-sparse and hence many of channel coefficients are zeros [14] [15]. Hence, the state-of-the-art algorithms do not exploit sparse channel structure information and there are some performance gain could be obtained if we can develop advanced adaptive channel estimation methods.

In this paper, we propose three sparse VSS-APSA algorithms by adopting three sparse constraint functions, i.e., zero-attracting (ZA) [16], reweighted zero attracting (RZA) [16] and reweighted $\ell_1$-norm (RL1) [17], to exploit channel sparsity as well as to mitigate impulsive interferences. Our contribution of this paper can be summarized as follows. First, cost function of zero-attracting VSS-APSA (ZA-VSS-APSA) is constructed and the corresponding update equation is derived. Second, reweighted zero-attracting VSS-APSA (RZA-VSS-APSA) and reweighted $\ell_1$-norm VSS-APSA (RL1-VSS-APSA) are developed as well. Third, strength of three sparse constraints is compared. At last, several representative simulation results are provided to verify the effectiveness of the proposed algorithms.

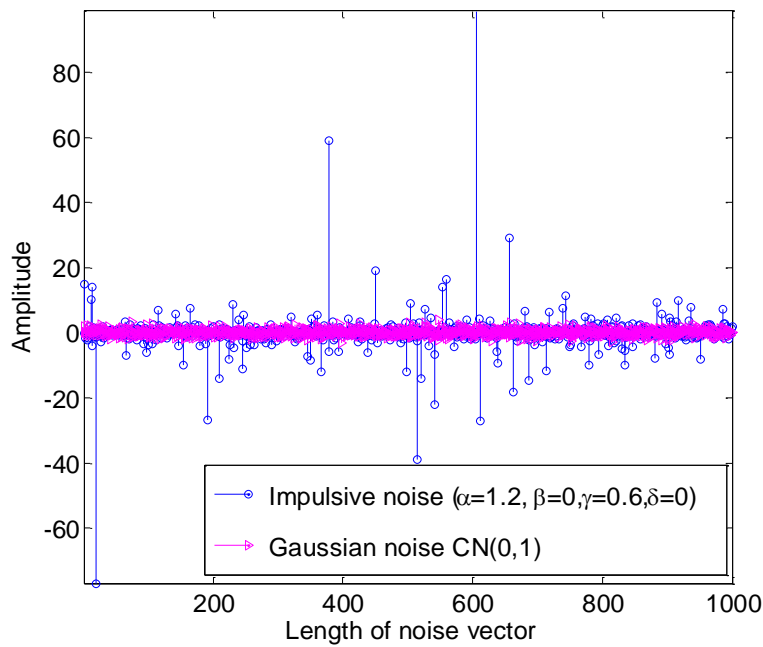

Fig. 1. Example of two noise models: impulsive noise vs. Gaussain noise.

The rest of the paper is organized as follows. Section 2 introduces an alpha-stable impulsive noise based sparse system model and find out the drawbacks of standard VSS-APSA. Based on sparse channel model, we propose three sparse VSS-APSA algorithms in Section 3. In Section 4, computer simulations are provided to validate the effectiveness of the propose algorithms. Finally, Section 5 concludes the paper and brings forward the future work.

## 2. System model and problem formulation

Let us consider an additive alpha-stable noise interference channel, which is modeled by the unknown $N$-length finite impulse response (FIR) vector $\mathbf{w} = [w_0, w_1, \ldots, w_{N-1}]^T$. The ideal received signal is expressed as

$$d(n) = \mathbf{x}^T(n)\mathbf{w} + z(n) \tag{1}$$

where $\mathbf{x}(n) = [x(n), x(n-1), \ldots, x(n-N+1)]^T$ is the input signal vector of the $N$ most recent input samples and $z(n)$ is a $\alpha$-stable noise. At the time $t$, the characteristic function of alpha-stable process $p(t)$ is defined as

$$p(t) = \exp\{j\delta t - \gamma |t|^\alpha [1 + j\beta \mathrm{sgn}(t) S(t, \alpha)]\} \tag{2}$$

where

$$S(t, \alpha) = \begin{cases} \tan(\alpha\pi/2), & if\ \alpha \neq 1 \\ (\pi/2)\log(t), & if\ \alpha = 1 \end{cases} \tag{3}$$

Here, $\alpha \in (0,2]$ denotes the characteristic exponent to measure the tail heaviness of the distribution, i.e., smaller $\alpha$ indicates heavier tail and vice versa. One can find that the Gaussian process is a special case of alpha-stable noise when $\alpha = 2$. $\gamma > 0$ represents the dispersive parameter to act a similar role to the variance of Gaussian distribution; $\beta \in [-1,1]$ denotes the symmetrical parameter which controls symmetry scenarios about its local parameter $\delta$. Throughout noise is symmetrical in the case of $\beta = 0$ as well as $\delta = 0$. The objective of the adaptive channel estimation is to perform adaptive estimate of $\widetilde{\mathbf{w}}(n)$ with limited complexity and memory given sequential observation $\{d(n), \mathbf{x}(n)\}$ in the presence of additive noise $z(n)$. We define the a prior output error vector, and the a posteriori output error vector as

$$e(n) = d(n) - \mathbf{x}^T(n)\widetilde{\mathbf{w}}(n) \tag{4}$$

$$e_p(n) = d(n) - \mathbf{x}^T(n)\widetilde{\mathbf{w}}(n+1) \tag{5}$$

where $\widetilde{\mathbf{w}}(n)$ is the estimator of $\mathbf{w}_0$ at iteration $n$. The standard VSS-APSA is derived by minimizing the $\ell_1$-norm of the a posteriori output vector with a constraint on the weight coefficients vectors, as

$$\underset{\widetilde{\mathbf{w}}(n+1)}{\text{minimize}} \, \|e_p(n)\|_1 \tag{6}$$

$$\text{subject to } \|\widetilde{\mathbf{w}}(n+1) - \widetilde{\mathbf{w}}(n)\|_2^2 \leq \mu^2$$

where $\mu$ is a parameter used to guarantee that the weight coefficient vectors do not change abruptly. This constrained minimization problem can be solved using the Lagrange multipliers with the constraint (6). The minimum disturbance $\varepsilon$ controls the convergence level of the algorithm and it shall be as small as possible. Using the method of Lagrange multipliers, the unconstrained cost function $G_s(n+1)$ can be obtained by combining (5) and (6),

$$G_s(n) = \|e_p(n)\|_1 + \beta(\|\widetilde{\mathbf{w}}(n+1) - \widetilde{\mathbf{w}}(n)\|_2^2 - \mu^2) \tag{7}$$

where $\beta$ is a Lagrange multiplier. The derivative of the cost function (7) with respect to the weight vector $\widetilde{\mathbf{w}}(n+1)$ is

$$\frac{\partial G_s(n)}{\partial \widetilde{\mathbf{w}}(n)} = -\text{sgn}\left(e_p(n)\right)\mathbf{x}(n) + 2\beta\left(\widetilde{\mathbf{w}}(n+1) - \widetilde{\mathbf{w}}(n)\right) \tag{8}$$

where $sgn(\cdot)$ denotes the sign function. Setting the derivative of $G_s(n+1)$ equal to zero, the updating equation of standard VSS-APSA is obtained as

$$\widetilde{\mathbf{w}}(n+1) = \widetilde{\mathbf{w}}(n) + \frac{1}{2\beta}\mathbf{x}(n)\text{sgn}\left(e_p(n)\right) \tag{9}$$

Substituting (9) into (8), one can get

$$\frac{1}{2\beta} = \frac{\mu}{\sqrt{\text{sgn}\left(e_p^T(n)\right)\mathbf{x}^T(n)\mathbf{x}(n)\text{sgn}\left(e_p(n)\right)}} \tag{10}$$

$$= \frac{\mu}{\|\mathbf{x}(n)\|_2}$$

with respect to the weight. Hence, substituting (10) into (9), the weight coefficient vector of the standard VSS-APSA is given recursively using the following update equation

$$\widetilde{\mathbf{w}}(n+1) = \widetilde{\mathbf{w}}(n) + \frac{\mu}{\|\mathbf{x}(n)\|_2 + \delta_0}\mathbf{x}(n)\text{sgn}\left(e_p(n)\right) \tag{11}$$

$$= \widetilde{\mathbf{w}}(n) + \mu(n)\mathbf{x}(n)\text{sgn}\left(e_p(n)\right),$$

where $\mu$ and $\mu(n)$ act as initial step-size (ISS) and variable step-size, respectively, and $\delta_0 > 0$ denotes a regularization parameter. One can find that the $\mu(n)$ in (11) depends on the input signal $\mathbf{x}(n)$ and $\delta_0$. Even though the above algorithm in (11) can mitigate the impulsive interference, sparse structure in channel could not be exploited for neglecting sparse constraint on update channel vectors.

### 3. Proposed sparse VSS-APSA algorithms

#### 3.1. Optimal sparse VSS-APSA with $\ell_0$-norm sparse constraint

To full take advantage of channel sparsity, optimal sparse constraint function (i.e., $\ell_0$-norm) is considered for sparse VSS-APSA algorithm for estimating channels in impulsive interference environments. With a constraint on the weight channel coefficients vectors, hence, the optimal sparse VSS-APSA is derived by minimizing *the affine combination* of $\ell_1$-norm a posteriori output vector and $\ell_0$-norm updating channel vector as

$$\underset{\widetilde{\mathbf{w}}(n+1)}{\text{minimize}} \ \|e_p(n)\|_1 + \lambda_0 \|\widetilde{\mathbf{w}}(n)\|_0 \tag{12}$$

$$\text{subject to } \|\widetilde{\mathbf{w}}(n+1) - \widetilde{\mathbf{w}}(n)\|_2^2 \leq \varepsilon^2$$

where $\|\cdot\|_0$ represents $\ell_0$-norm function and $\lambda_0$ denotes the regularization parameter to trade off the instantaneous estimation error and $\ell_0$-norm sparse penalty of $\widetilde{\mathbf{w}}(n+1)$. In the perspective of mathematical theory, adopting the $\ell_0$-norm as for sparse constraint function could exploit maximal sparse structure information in channels. However, by solving the $\ell_0$-norm is a NP-hard (non-deterministic polynomial-time hard) problem [18]. Hence, it is necessary to replace it with approximate sparse constraints so that (12) can be solvable. In the subsequent, we propose three alternative sparse adaptive filtering algorithms, i.e., ZA-VSS-APSA, RZA-VSS-APSA as well as RL1-

VSS-APSA, to exploit the channel sparsity as well as to mitigate the impulsive interferences simultaneously.

### 3.2. ZA-VSS-APSA

According to (12), one can replace the $\ell_0$-norm sparse constraint with $\ell_1$-norm function [19] and then construct the cost function of ZA-VSS-APSA as follows:

$$G_{za}(n) = \|e_p(n)\|_1 + \lambda_{za}\|\widetilde{\mathbf{w}}(n)\|_1 + \beta(\|\widetilde{\mathbf{w}}(n+1) - \widetilde{\mathbf{w}}(n)\|_2^2 - \varepsilon^2) \tag{13}$$

where $\lambda_{za}$ denotes a regularization parameter to balance between estimation error and $\ell_1$-norm sparse constraint function of the $\widetilde{\mathbf{w}}(n)$. The derivative of the cost function (13) with respect to the weight vector $\widetilde{\mathbf{w}}(n)$ is

$$\frac{\partial G_{za}(n)}{\partial \widetilde{\mathbf{w}}(n)} = -\mathrm{sgn}\left(e_p(n)\right)\mathbf{x}(n) + \lambda_{za}\mathrm{sgn}(\widetilde{\mathbf{w}}(n)) + 2\beta(\widetilde{\mathbf{w}}(n+1) - \widetilde{\mathbf{w}}(n)) \tag{14}$$

By means of (14), setting the derivative of $G_{za}(n)$ equal to zero, the update equation of ZA-VSS-APSA based sparse channel estimation method can be derived as

$$\widetilde{\mathbf{w}}(n+1) = \widetilde{\mathbf{w}}(n) + \underbrace{\mu(n)\mathbf{x}(n)\mathrm{sgn}\left(e_p(n)\right)}_{\text{to mitigate impulsive noise}} - \underbrace{\lambda_{za}/2\,\mathrm{sgn}(\widetilde{\mathbf{w}}(n))}_{\text{to exploit sparsity}} \tag{15}$$

To analysis the update stability of proposed ZA-VSS-APSA, MSD performance is analyzed as follows.

### 3.3. RZA-VSS-APSA

It was well known that more strong sparse constraint could exploit sparsity more efficient [17]. This principle implies that channel estimation performance could be improved by using more efficient sparse approximation function even if in the presence of impulsive interferences. Hence, the cost function of RZA-VSS-APSA is written as

$$G_{za}(n) = \|e_p(n)\|_1 + \lambda_{rza}\sum_{i=0}^{N-1}\log(1 + \varepsilon_{rza}|\widetilde{w}_i(n)|) + \beta(\|\widetilde{\mathbf{w}}(n+1) - \widetilde{\mathbf{w}}(n)\|_2^2 - \varepsilon^2) \tag{16}$$

where $\lambda_{rza} > 0$ is *a regularization parameter* to balance the estimation error and sparsity of $\sum_{i=0}^{N-1}\log(1 + \varepsilon_{rza}\widetilde{w}_i(n))$. Likewise, the corresponding update equation can be derived as

$$\widetilde{\mathbf{w}}(n+1) = \widetilde{\mathbf{w}}(n) + \underbrace{\mu(n)\mathbf{x}(n)\text{sgn}\left(e_p(n)\right)}_{\text{to mitigate impulsive noise}} - \underbrace{\frac{\lambda_{rza}}{2} \cdot \frac{\text{sgn}(\widetilde{\mathbf{w}}(n))}{1+\varepsilon_{rza}|\widetilde{\mathbf{w}}(n)|}}_{\text{to exploit sparsity}} \quad (17)$$

where reweighted factor is set as $\varepsilon_{rza} = 20$ [20] to exploit channel sparsity efficiently. In the second term of (16), please notice that estimated channel coefficients $|\widetilde{w}_i(n)|, i = 0,1,\dots, N-1$ are replaced by zeroes in high probability if under the hard threshold $1/\varepsilon_{rza}$. Hence, one can find that RZA-VSS-APSA can exploit sparsity and mitigate noise interference simultaneously.

### 3.4. RL1-VSS-APSA

Beside the RZA-type algorithm, RL1 minimization for adaptive sparse channel estimation has a better performance than L1 minimization that is usually employed in compressive sensing [17]. It is due to the fact that a properly reweighted $\ell_1$-norm approximates the $\ell_0$-norm, which actually needs to be minimized, better than $\ell_1$-norm. Hence, one approach to enforce the sparsity of the solution for the sparse VSS-APSA algorithms is to introduce the RL1 penalty term in third cost function as RL1-VSS-APSA which considers a penalty term proportional to the RL1 of the coefficient vector. The corresponding cost function can be written as

$$G_{rl1}(n) = \|e_p(n)\|_1 + \lambda_{rl1}\|\boldsymbol{f}(n)\widetilde{\mathbf{w}}(n)\|_1 + \beta(\|\widetilde{\mathbf{w}}(n+1) - \widetilde{\mathbf{w}}(n)\|_2^2 - \varepsilon^2) \quad (18)$$

where $\lambda_{rl1}$ is the weight associated with the penalty term and elements of the $1 \times N$ row vector $\boldsymbol{f}(n)$ are set to

$$[\boldsymbol{f}(n)]_i = \frac{1}{\delta_{rl1} + |[\widetilde{\mathbf{w}}(n)]_i|}, i = 0,1,\dots,N-1 \quad (19)$$

where $\delta_{rl1}$ being some positive number and hence $[\boldsymbol{f}(n)]_i > 0$ for $i = 0,1,\dots,N-1$. The update equation can be derived by differentiating (18) with respect to the FIR channel vector $\widetilde{\mathbf{w}}(n)$. Then, the resulting update equation is:

$$\widetilde{\mathbf{w}}(n+1) = \widetilde{\mathbf{w}}(n) + \underbrace{\mu(n)\mathbf{x}(n)\text{sgn}\left(e_p(n)\right)}_{\text{to mitigate impulsive noise}} - \underbrace{\frac{\lambda_{rl1}}{2} \cdot \frac{\text{sgn}(\widetilde{\mathbf{w}}(n))}{\delta_{rl1}+\widetilde{w}(n-1)}}_{\text{to exploit sparsity}} \quad (20)$$

Please notice that in Eq. (19), since $\text{sgn}(\boldsymbol{f}(n)) = \mathbf{1}_{1\times N}$, hence one can get $\text{sgn}(\boldsymbol{f}(n)\widetilde{\mathbf{w}}(n)) = \text{sgn}(\widetilde{\mathbf{w}}(n))$. Note that although the weight vector $\widetilde{\mathbf{w}}(n)$ changes in every stage of this sparsity-aware RL1-VSS-APSA algorithm, it does not depend on $\widetilde{\mathbf{w}}(n)$, and the cost function $G_{rl1}(n)$ is convex. Therefore, the RL1 penalized RL1-VSS-APSA is guaranteed to converge to the global minimization under some conditions.

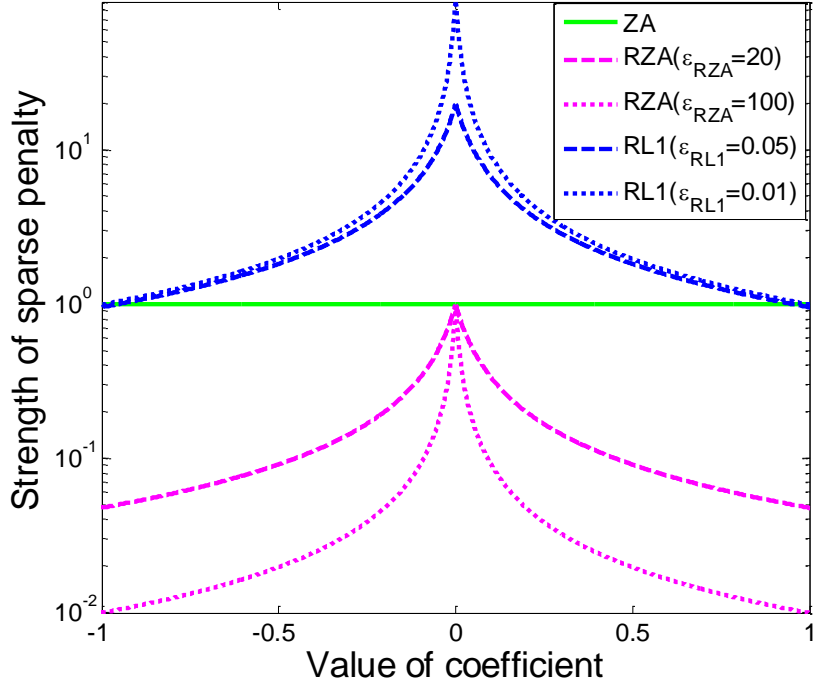

Fig. 2. Sparse penalty strength in different sparse constraint functions.

### 3.5. Evaluation the strength of of sparse constraints

To evaluate the sparse penalty strength of ZA, RZA and RL1, we define above three sparse penalty functions as follows:

$$\zeta_{za} = \text{sgn}(\widetilde{\mathbf{w}}(n)) \tag{21}$$

$$\zeta_{rza} = \frac{\text{sgn}(\widetilde{\mathbf{w}}(n))}{1 + \varepsilon_{rza}|\widetilde{\mathbf{w}}(n)|} \tag{22}$$

$$\zeta_{rl1} = \frac{\text{sgn}(\widetilde{\mathbf{w}}(n))}{\delta_{rl1} + |\widetilde{\mathbf{w}}(n-1)|} \tag{23}$$

where channel coefficients in $\mathbf{w}$ are assumed in range $[-1,1]$. Considering above three sparse constraints in Eqs. (21)~(23), their sparse penalty strength curves are depicted in Fig. 2. One can find that ZA utilizes uniform sparse penalty to all channel coefficients in the range of $[-1,1]$ and hence it is not efficient to exploit channel sparsity. Unlike the ZA (21), both RZA (22) and RL1 (23) make use of adaptively sparse penalty on different channel coefficients, i.e., stronger sparse penalty on zero/approximate zero coefficients and weaker sparse penalty on significant coefficients. In addition, one can also find that RL1 (23) utilizes stronger sparse penalty than RZA (22) as shown in Fig. 2. Hence, RL1-VSS-APSA can exploit more sparse information than both ZA-VSS-APSA and RZA-VSS-APSA on adaptive sparse channel estimation in the presence of impulsive interferences.

## 4. Computer simulations and discussions

In this section, the proposed channel estimation methods are evaluated in different impulsive environments. For achieving average performance, $M = 1000$ independent Monte-Carlo runs are adopted. The simulation setup is configured according to typical broadband wireless communication system in Japan. The signal bandwidth is $60MHz$ located at the central radio frequency of 2.1GHz. The maximum delay spread of $1.06\mu s$. Hence, the maximum length of channel vector $\mathbf{w}$ is $N = 128$ and its number of dominant taps is set to $K \in \{4,8,12\}$. To validate the effectiveness of the proposed methods, average mean square error (MSE) standard is adopted. Channel estimators are evaluated by average MSE which is defined by

$$Average\ MSE\{\widetilde{\mathbf{w}}(n)\} := 10\log_{10}\frac{1}{M}\sum_{m=1}^{M}\frac{\|\widetilde{\mathbf{w}}^m(n) - \mathbf{w}\|_2^2}{\|\mathbf{w}\|_2^2} \qquad (24)$$

where $\mathbf{w}$ and $\widetilde{\mathbf{w}}(n)$ are the actual signal vector and reconstruction vector, respectively. The results are averaged over 1000 independent Monte-Carlo runs. Each dominant channel tap follows random Gaussian distribution as $\mathcal{CN}(0,\sigma_w^2)$ which is subject to $E\{\|\mathbf{w}\|_2^2\} = 1$ and their positions are randomly decided within the $\mathbf{w}$. The received SNR is defined as $10\log(E_s/\sigma_n^2)$, where $E_s$ is the received power of the pseudo-random noise (PN)-sequence for training signal. In addition, to achieve better steady-state estimation performance, reweighted factor of RZA-VSS-APSA is set as $\varepsilon_{rza} = 20$ [20][21]. Threshold parameter of RL1-VSS-APSA is set as $\delta_{rl1} = 0.01$ [22]. Detailed parameters for computer simulation are given in Tab. 1.

**Tab. 1. Simulation parameters.**

| Parameters | Values |
|---|---|
| Training signal | Pseudo-random Gaussian sequence |
| alpha-stable noise distribution | $\alpha \in \{1.2, 1.8\}, \beta = 0,$ $\gamma \in \{0.6, 1.2\}, \delta = 0$ |
| Channel length | $N = 128$ |
| No. of nonzero coefficients | $K \in \{4,8,12\}$ |
| Distribution of nonzero coefficient | Random Gaussian $\mathcal{CN}(0,1)$ |
| Regularization parameter for VSS-APSA | $\delta_0 = 10^{-6}$ |
| Received SNR for channel estimation | $\{-5\text{dB}, 5\text{dB}\}$ |
| Initial step-size | $\mu = 0.1$ |
| Regularization parameters for sparse penalties | $\lambda_{za} = 0.0004, \lambda_{rza} = 0.004$ and $\lambda_{rl1} = 0.0001$ |
| Reweight factor of RZA-VSS-APSA | $\varepsilon_{rza} = 20$ |
| Threshold of the RL1-VSS-APSA | $\delta_{rl1} = 0.01$ |

In the first example, average MSE performances of the proposed methods are evaluated for $K \in \{4,8,12\}$ in Figs. 3-5 under two SNR regimes ($-5\text{dB}$ and $5\text{dB}$). To confirm the effectiveness of the three proposed methods, they are compared with standard VSS-APSA [23]. One can find the proposed sparse VSS-APSA algorithms always achieve better performance with respect to average MSE. In the case of impulsive noise ($\alpha = 1.2, \beta = 0, \gamma = 0.6, \delta = 0$), our proposed algorithm can get at least 10dB performance gain according to the Figs. 3-4 for very sparse channel (e.g., $K = 4$ and 8). While in low SNR regime (e.g. $\text{SNR} = -5\text{dB}$), Fig. 5 shows that the proposed algorithms can still get 3dB performance gain even if $K = 12$. Hence, the effectiveness of the proposed algorithms is confirmed in the case of different sparse channels under different SNR regimes.

In the second example, the proposed methods are evaluated in different impulsive interferences. It is well known that robust performance of proposed algorithms may depend highly on different impulsive interferences. Here, three kinds of impulsive interferences are considered in Figs. 6~8 to evaluate the average MSE of the proposed algorithms in the case of SNR = 5dB and $K = 8$. One can find that our proposed algorithms can achieve a lot performance gains (at least 10dB) and can robust mitigate different impulsive interferences even if in the case of very strong interference environments, e.g., α = 1.2 and γ = 1.2.

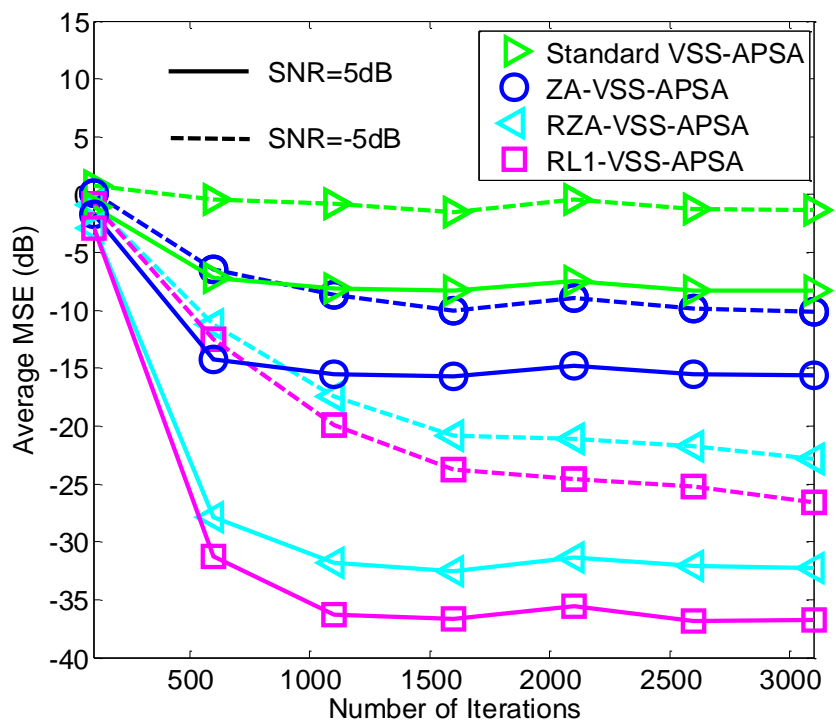

Fig. 3. Avergae MSE comparsions v.s. number of iterations ($K$=4, $\alpha = 1.2, \beta = 0, \gamma = 0.6, \delta = 0$).

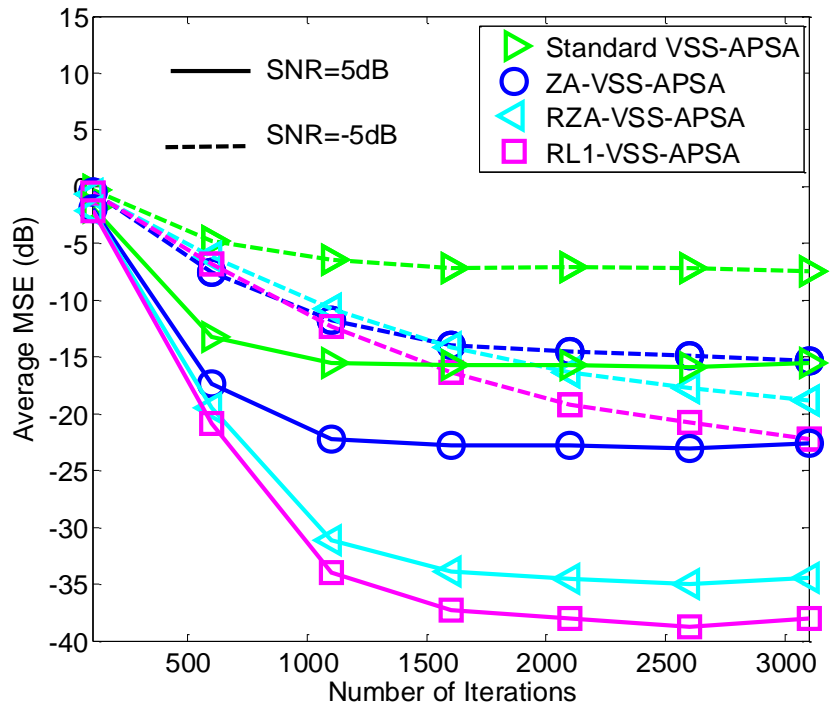

Fig. 4. Avergae MSE v.s. number of iterations (K=8, $\alpha = 1.2, \beta = 0, \gamma = 0.6, \delta = 0$).

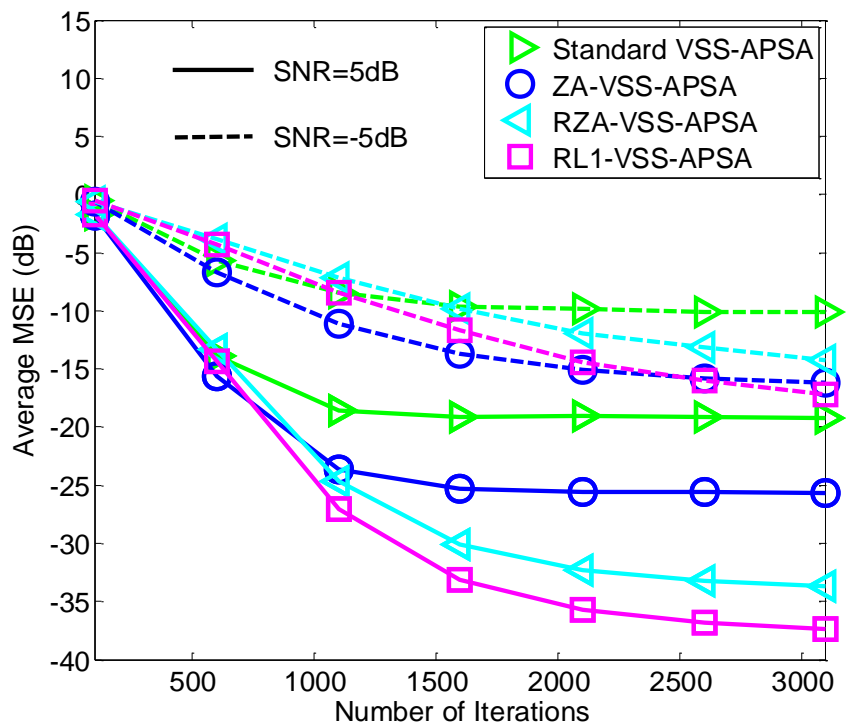

Fig. 5. Avergae MSE v.s. number of iterations ($K$=12, $\alpha = 1.2, \beta = 0, \gamma = 0.6, \delta = 0$).

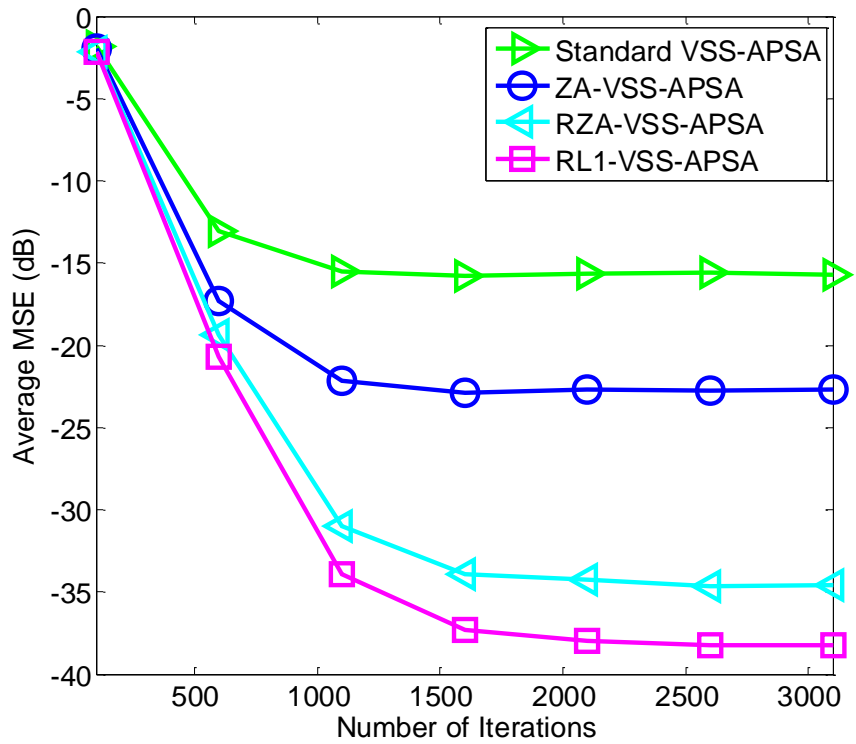

Fig. 6. Avergae MSE v.s. number of iterations ($\alpha = 1.2, \beta = 0, \gamma = 0.6, \delta = 0$).

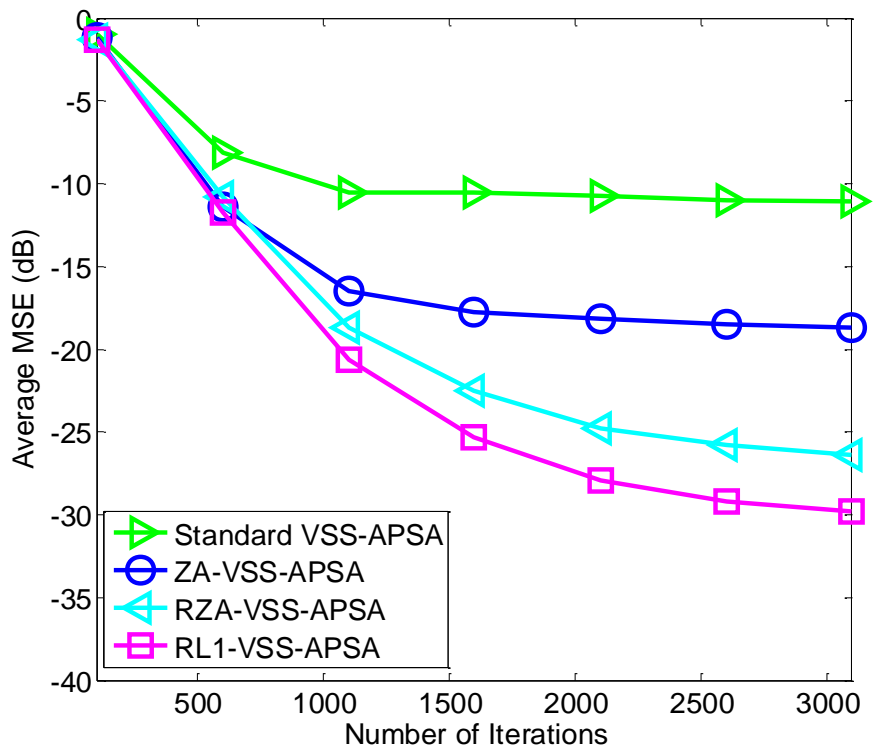

Fig. 7. Avergae MSE v.s. number of iterations ($\alpha = 1.2, \beta = 0, \gamma = 1.2, \delta = 0$).

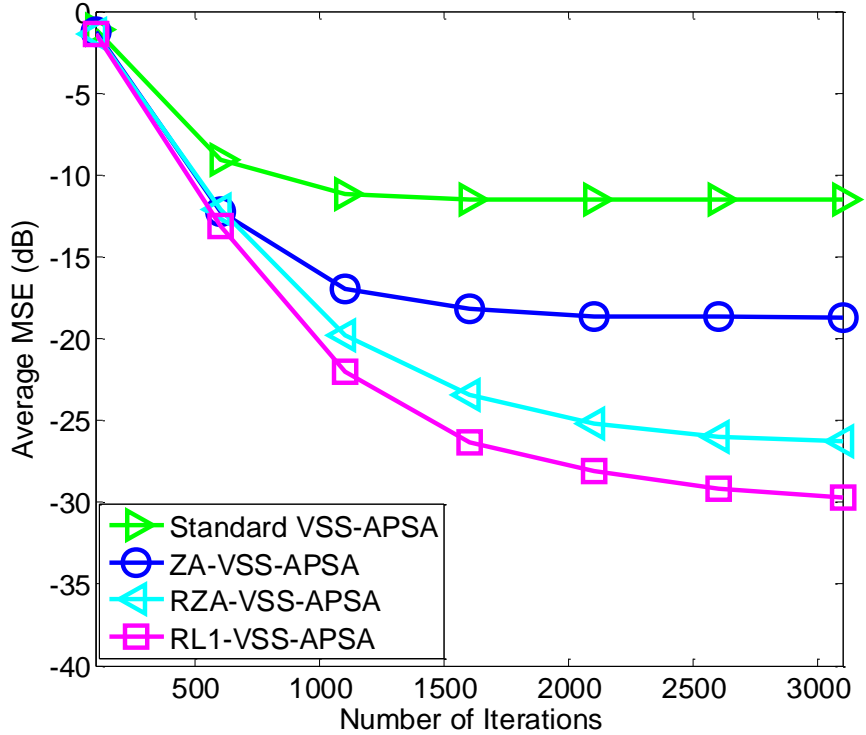

Fig. 8. Avergae MSE v.s. number of iterations ($\alpha = 1.8, \beta = 0, \gamma = 1.2, \delta = 0$).

## 5. Conclusions and future work

Conventional adaptive sparse channel estimation algorithms were proposed to reconstruct channel as well as to exploit channel sparsity under Gaussian noise environments. In the case of impulsive interference, the previous algorithms are vulnerable to deteriorate. This paper proposed three robust sparse VSS-APSA algorithms for mitigating the impulsive noise interferences as well as exploiting channel sparisty. Both theoretical analysis and computer simulation verified the performance gain in different impulsive levels. In future work, our proposed methods will be tested in different communications systems, such as underwater acoustic systems as well as power-line communication systems.

## Acknowledgment

This work was supported in part by Japan Society for the Promotion of Science (JSPS) research activity start-up research grant (No. 26889050) as well as the National Natural Science Foundation of China grants (No. 61401069, No. 61261048, No. 61201273).

## References

[1]  F. Adachi, D. Garg, S. Takaoka, and K. Takeda, "Broadband CDMA techniques," *IEEE Wirel. Commun.*, vol. 12, no. April, pp. 8–18, 2005.

[2]  B. D. Raychaudhuri and N. B. Mandayam, "Frontiers of wireless and mobile communications," *Proc. IEEE*, vol. 100, no. 4, 2012.

[3]  L. Dai, Z. Wang, and Y. Zhixing, "Next-generation digital television terrestrial broadcasting systems: Key technologies and research trends," *IEEE Commun. Mag.*, vol. 50, no. 6, pp. 150–158, 2012.

[4]  L. Dai, Z. Wang, and Z. Yang, "Compressive sensing based time domain synchronous OFDM transmission for vehicular communications," *IEEE J. Sel. Areas Commun.*, vol. 31, no. 9, pp. 460–469, 2013.

[5]  L. Dai, J. Wang, S. Member, Z. Wang, P. Tsia, and M. Moonen, "Spectrum-and energy-efficient OFDM based on simultaneous multi-channel reconstruction," vol. 61, no. 23, pp. 6047–6059, 2013.

[6]  Z. Gao, L. Dai, and Z. Wang, "Structured compressive sensing based superimposed pilot design in downlink large-scale MIMO systems," *Electron. Lett.*, vol. 50, no. 12, pp. 896–898, 2014.

[7]  L. Dai, Z. Wang, and Z. Yang, "Spectrally efficient time-frequency training OFDM for mobile large-scale MIMO systems," *IEEE J. Sel. Areas Commun.*, vol. 31, no. 2, pp. 251–263, 2013.

[8]  S. Haykin, *Adaptive filter theory*, NJ: Prentice–Hall, 1996.

[9]  M. Shao and C. L. Nikias, "Signal processing with fractional lower order moments: stable processes and their applications," *Proc. IEEE*, vol. 81, no. 9210747, pp. 986–1010, 1993.

[10] J. Lin, S. Member, M. Nassar, and B. L. Evans, "Impulsive noise mitigation in powerline communications using sparse Bayesian learning," vol. 31, no. 7, pp. 1172–1183, 2013.

[11] Y.-P. Li, T.-S. Lee, and B.-F. Wu, "A variable step-size sign algorithm for channel estimation," *Signal Processing*, vol. 102, pp. 304–312, Sep. 2014.

[12] T. Shao, Y. Zheng, and J. Benesty, "An affine projection sign algorithm robust against impulsive interferences," *Signal Process. Lett. IEEE*, vol. 17, no. 4, pp. 327–330, 2010.

[13] J. Yoo, J. Shin, and P. Park, "Variable step-size affine projection aign algorithm," *IEEE Trans. Circuits Syst. Express Briefs*, vol. 61, no. 4, pp. 274–278, 2014.

[14] L. Dai, G. Gui, Z. Wang, Z. Yang, and F. Adachi, "Reliable and energy-efficient OFDM based on structured compressive sensing," in *IEEE International Conference on Communications (ICC), Sydney, Australia, 10-14 June 2014*, 2014, pp. 1–6.

[15] G. Gui, W. Peng, and F. Adachi, "Sub-Nyquist rate ADC sampling-based compressive channel estimation," *Wirel. Commun. Mob. Comput.*, vol. 13, no. 18, pp. 1–10, 2013.

[16] Y. Chen, Y. Gu, A. O. Hero III, A. Hero, and A. O. H. Iii, "Sparse LMS for system identification," in *IEEE International Conference on Acoustics, Speech and Signal Processing, April 19-24, 2009,Taipei, Taiwan*, 2009, no. 3, pp. 3125–3128.

[17] E. J. Candes, M. B. Wakin, and S. P. Boyd, "Enhancing sparsity by reweighted L1 minimization," *J. Fourier Anal. Appl.*, vol. 14, no. 5–6, pp. 877–905, 2008.

[18] D. L. Donoho and X. Huo, "Uncertainty principles and ideal atomic decomposition," *IEEE Transcations Inf. Theory*, vol. 47, no. 7, pp. 2845–2862, 2001.

[19] D. L. L. Donoho, "Compressed sensing," *IEEE Trans. Inf. Theory*, vol. 52, no. 4, pp. 1289–1306, 2006.

[20] G. Gui, A. Mehbodniya, and F. Adachi, "Least mean square/fourth algorithm for adaptive sparse channel estimation," in *IEEE International Symposium on Personal, Indoor and Mobile Radio Communications (PIMRC), Sept. 8-11, 2013, London, UK*, 2013, pp. 1–5.

[21] G. Gui, L. Dai, S. Kumagai, and F. Adachi, "Variable earns profit: Improved adaptive channel estimation using sparse VSS-NLMS algorithms," in *in IEEE International Conference on Communications (ICC), Sydney, Australia, 10-14 June 2014*, 2014, pp. 1–5.

[22] O. Taheri and S. A. Vorobyov, "Sparse channel estimation with Lp-norm and reweighted L1-norm penalized least mean squares," in *IEEE International Conference on Acoustics, Speech and Signal Processing (ICASSP), Prague, Czech Republic, 22-27 May*, 2011, pp. 2864–2867.

[23] H. Shin, A. H. Sayed, and W. Song, "Variable step-size NLMS and affine projection algorithms," *IEEE Signal Process. Lett.*, vol. 11, no. 2, pp. 132–135, 2004.